\documentclass[aoas]{imsart}

\RequirePackage{amsthm,amsmath,amsfonts,amssymb}
\RequirePackage[authoryear]{natbib}
\RequirePackage[colorlinks,citecolor=blue,urlcolor=blue]{hyperref}
\RequirePackage{graphicx}
\usepackage{booktabs}
\usepackage{float}
\usepackage{pdflscape}
\usepackage{makecell}

\startlocaldefs
\theoremstyle{plain}

\theoremstyle{remark}

\newcommand{\xl}[1]{\textcolor{red}{#1}}

\endlocaldefs

\begin{document}

\begin{frontmatter}
\title{Clustering of functional data prone to complex heteroscedastic measurement error}
\runtitle{Clustering of functional data prone to complex heteroscedastic measurement error}

\begin{aug}
\author[A]{\fnms{Andi}~\snm{Mai}\orcid{0000-0002-xxxx-xxxx}},
\author[B]{\fnms{Lan}~\snm{Xue}\orcid{0000-0002-7764-8564}},
\author[A]{\fnms{Roger}~\snm{Zoh}\orcid{0000-0002-8066-1153}},
\author[A]{\fnms{Carmen}~\snm{Tekwe}\orcid{0000-0002-1857-2416}}
\address[A]{Department of Epidemiology and Biostatistics,
Indiana University - Bloomington}

\address[B]{Department of Statistics,
Oregon State University}
\end{aug}

\begin{abstract}
Several factors make clustering of functional data challenging, including the infinite dimensional space to which observations belong and the lack of a defined probability density function for the functional random variable. To overcome these barriers, researchers either assume that observations belong to a finite dimensional space spanned by basis functions or apply nonparametric smoothing methods to the functions prior to clustering. Although extensive literature describes clustering methods for functional data, few studies have explored the clustering of error–prone functional data. In this work, we consider clustering methods for functional data prone to complex, heteroscedastic measurement errors and propose a two-stage-based approach for clustering. Under the first stage, clustered mixed effects models are applied to adjust for measurement error bias, followed by cluster analysis of the measurement error–adjusted curves in the second stage. The cluster analysis can be performed using readily available methods for cluster analysis such as K-means and mclust. Through simulations, we investigate how varying sample sizes, the magnitude of measurement error and the correlation structure associated with the measurement errors influence the clustering of the error prone data. Our results indicate that failing to account for measurement errors and the correlation structures associated with frequently collected functional data reduces the accuracy of identifying the true latent groups or clusters. The developed methods are applied to two data sets, a school-based study of energy expenditure among elementary school-aged children in Texas and data from the National Health and Nutritional Examination Survey on participants’ physical activity monitored by wearable devices at frequent intervals. 
\end{abstract}


\end{frontmatter}

\section{Introduction}

Functional data analysis involves the examination of data that appear in forms such as functions, images, or other complex objects [\cite{wang2016functional,katina2021functional,srivastava2016motivation}]. Distinct from traditional multivariate data analysis, functional data are characterized by their inherently infinite-dimensional nature [\cite{ullah2013applications,wang2024functional,lin2018mixture}]. The term functional data analysis was first coined by Ramsay [\cite{ramsay1982data,ramsay1988monotone,ramsay1991some}] where  justifications were provided for introducing and proposing functional data analysis as an analytical tool for data that appear as functions. Functional data analysis provides a smooth, comprehensive representation of data, facilitating more realistic modeling and the identification of trends across the entire domain, rather than focusing on discrete points[\cite{kokoszka2017introduction,levitin2007introduction,rossi2005representation,ferraty2006nonparametric}]. By considering temporal and spatial correlations, functional data analysis enhances model accuracy and mitigates issues such as overfitting and underestimated variability[\cite{li2014functional,ruggieri2018comparing}]. However, analyzing functional data is challenging due to its high dimensionality, irregularity, complex covariance structures, measurement errors, and the difficulty with model interpretations[\cite{james2009functional,wang2016functional,ullah2013applications,cuevas2014partial,fan2014challenges,tekwe2019instrumental}]. 

Several methods have been developed for the analysis of functional data including functional regression[\cite{morris2015functional,greven2017general,goldsmith2011penalized,cai2006prediction,crainiceanu2009generalized}], clustering, and classification of functional data[\cite{chamroukhi2019model,rossi2006support,leng2006classification,delaigle2012achieving,baillo2010classification}]. Functional regression extends classical regression methods to situations where either the predictors, the responses, or both are functions[\cite{morris2015functional}]. 
Functional regression models can capture time or space varying dynamics [\cite{kowal2021dynamic,cai2000functional,zhu2018robust}]. Analysis of functional data typically involves capturing important patterns in the data by smoothing the data that are assumed to be observed discrete realizations of an underlying latent curve[\cite{ramsay2009introduction}]. Clustering and classification methods for functional data group or classify the functions or curves into distinct categories based on their shapes, behaviors, or patterns[\cite{jacques2014functional,tarpey2003clustering}]. Approaches for classifying and clustering of functional data has gathered considerable attention in recent literature, leading to the development of numerous methodologies that can be broadly classified into several key categories[\cite{zhang2023review,chamroukhi2019model,ferreira2009comparison}] such as regression-based clustering, distance-based clustering, transformation-based clustering, non-parametric based clustering, Bayesian-based methods and other emerging techniques. 


Regression-based clustering encompasses a range of methods designed to leverage regression techniques for clustering functional data. Examples of regression-based clustering methods include functional latent mixture models[\cite{bouveyron2011model}], functional Gaussian mixture models[\cite{centofanti2023sparse,nguyen2016faster}], functional linear regression models[\cite{goia2010functional}], parametric mixture models[\cite{jacques2014model,jacques2013funclust}], and functional multiplicative models[\cite{li2011identifying}]. Moreover, mixed-effect models [\cite{misumi2019multivariate, ma2008penalized}] provide a robust framework for analyzing functional data by accounting for both fixed and random effects. When combined with wavelet decomposition techniques [\cite{giacofci2013wavelet, antoniadis2013clustering}], this approach enhances the ability to interpret complex patterns within both fixed and random components, enabling more detailed insights into functional data.

Another approach to clustering of functional data are based on differentials where clusters are defined based on the means and modes of variation between clusters[\cite{chiou2007functional}], or the covariances between two populations[\cite{ieva2016covariance}]. Methods based on the Mahalanobis distance in Hilbert spaces[\cite{martino2019k,galeano2015mahalanobis}] have also been used for clustering and classification of functional data. Additionally, distance-based two-layer partition clustering methods[\cite{qu2023robust,singh2017two}] have  been developed for clustering of functional data. Other differential-based methods for clustering of functional are based on the variations in phase and amplitude of the data  [\cite{park2017clustering,fu2019clustering,slaets2012phase}].

Transformed-based clustering includes techniques such as defining functions or curves by shifts and rotations as equivalent classes, employing rank-based transformations, and extracting splitting information from trajectories or their derivatives [\cite{guo2020robust,lim2019functional,justel2016sequential,pulido2021functional}]. Additionally, semi-parametric mixed normal transformation models and time-shift clustering algorithms further enhance the efficacy of transformed-based clustering [\cite{zhong2021cluster,wei2023novel}].

Non-parametric based clustering offers flexible and adaptive techniques for clustering functional data without assuming specific parametric forms [\cite{galvani2021funcc,hael2023quantile,song2023multi,miller2008nonparametric,boulle2012functional}]. Some common methods for non-parametric-based methods for clustering include deterministic iterative procedures[\cite{galvani2021funcc}], non-parametric quantile-based models and multivariate functional principal component scores[\cite{hael2023quantile,song2023multi}]. Further, similarity measures combined with agglomerative clustering algorithms[\cite{miller2008nonparametric}], and piece-wise constant density estimation[\cite{boulle2012functional}] contribute to the robustness and adaptability of non-parametric clustering methods.

Bayesian methods are increasingly applied to clustering of functional data by incorporating prior knowledge in the clustering [\cite{zoh2024bayesian,ray2006functional,rigon2023enriched,zhang2014joint,xian2022clustering,xian2024clustering,petrone2009hybrid}]. Some examples of Bayesian-based methods include include non-parametric Bayes wavelet models[\cite{ray2006functional}], Dirichlet mixture models[\cite{rigon2023enriched,zhang2014joint,petrone2009hybrid}], and variational Bayes algorithms[\cite{xian2022clustering,xian2024clustering}]. Zoh et al. [\cite{zoh2024bayesian}] introduced a non-parametric Bayesian scalar-on-function regression (SoFR) method that corrects measurement errors using an instrumental variable with a time-varying biasing factor, relaxing the assumption of white noise in measurement errors in SoFR models.

Other approaches such as sparse functional clustering, binary trees through recursive splitting of observations, and approximating self-organizing map algorithms have been developed for clustering of functional data   [\cite{vitelli2023novel,golovkine2022clustering,rossi2004clustering}]. While methods for clustering functional data are well established, less work has been considered in the literature to address the clustering of functional data in the presence of complex heteroscedastic measurement errors. Though some prior studies on clustering functional data have mentioned measurement errors, these methods assumed that the error terms were random independent noise[\cite{james2003clustering,rossi2004clustering,ma2006data}].

It is essential to account for measurement error when clustering error-prone functional data to ensure accurate and reliable results [\cite{fuller2009measurement}]. Measurement error is defined as the difference between the observed measurement of a particular quantity and its true, intrinsic value[\cite{bound2001measurement,fuller2009measurement}]. Repeated measurements of an identical quantity on a subject are generally expected to produce varying results[\cite{matthews1990analysis,oldham1962note}]. This variability can be attributed to natural variation within the subject, variations in the methodology of measurement, or a combination pf the two[\cite{bland1996measurement}]. While many data applications contain measurement errors, one of the strengths of functional data analysis is its adeptness in handling measurement errors, facilitated by the use of repeated measurements for each subject [\cite{wang2016functional}].

Several methods have been developed to cluster error-prone data, including approaches for scalar data [\cite{zhang2020model}] and for functional data [\cite{lin2000nonparametric}]. MCLUST is a model-based clustering approach that uses finite Gaussian mixture models to classify data into clusters using statistical model selection criteria, such as the Bayesian Information Criterion (BIC) [\cite{fraley2012mclust}]. Zhang et al. extended this framework to handle measurement error in scalar data, developing the MCLUST-ME model, which incorporates Gaussian finite mixture modeling to correct for measurement error under both known and estimable error distributions [\cite{zhang2020model}]. However, the lack of closed-form solutions for parameter estimation in MCLUST-ME leads to substantial computational challenges. These challenges become even more pronounced for multivariate or functional data, where the complexity of the model increases exponentially. Lin et al. proposed nonparametric local polynomial kernel regression methods to analyze clustered data settings where each cluster has multiple observations, extending applications to cases where predictors are measured accurately and where they are measured with error. To address measurement error in predictors, they applied the simulation-extrapolation method, enhancing nonparametric regression techniques to handle repeated, error-prone measurements within clusters [\cite{lin2000nonparametric}]. 

Measurement error can substantially bias clustering results, leading to misleading interpretations and conclusions about the underlying structures within the data[\cite{pankowska2020effect,kumar2007clustering,cooper1988effect}]. Since most existing methods for clustering of functional data do not explicitly account for measurement error, addressing this gap is crucial for improving the reliability and validity of clustering outcomes, especially in applications where precision is paramount. Wearable monitors are widely used for global surveillance of physical activity, aiding in understanding the distribution and extent of these behaviors[\cite{matthews2012best,bassett2012device}]. They also contribute to clarifying the relationship between physical activity and health outcomes[\cite{kraus2019daily,jakicic2019physical,strain2020wearable}]. Research indicates that quantifying the effects of measurement errors on physical activity estimates is essential, as factors such as device-related errors such as sensor accuracy and calibration issues; user-related errors including non-compliance and improper device usage; and variations in activity type or environmental conditions can all contribute to inaccuracies in measurements[\cite{ferrari2007role,fuller2020reliability,ward2004analysis,bassett2000validity,seneviratne2017survey,schukat2016unintended}].

Researchers are keen to understand how clusters of individuals with varying levels of physical activity compare. This article aims to bridge this gap by introducing a two stage-based method using point-wise mixed effect models with clustering. The method first corrects for biases in the error prone curves in the first stage and the second stage involves clustering procedures. Our developed approach can handle functional data with heteroscedastic measurement error, offering a straightforward interpretation of the results and demonstrating superior performance, particularly in small sample sizes. The remainder of this article is organized as follows: Section 2 presents the point-wise clustered mixed effect model with measurement error (CMEM-ME). We provide descriptions of the simulation studies in Section 3, and Section 4 provides applications of the methods to two real life datasets. Finally, we provide some concluding remarks and discuss some future research directions in Section 5.

\section{Method}

\subsection{Point-wise clustered mixed effect model with measurement error (PCME-ME)} 
To address the computational challenges in handling large, complex longitudinal data, \cite{cui2022fast} proposed Fast Univariate Inference (FUI) that fits univariate pointwise mixed-effects models. We propose a clustered mixed effect model with measurement error (CMEM-ME) for correcting error-prone functional data. The pointwise clustered mixed effect model is considered by fitting a separate model for each time point across all replicates at each time \(t\). The function \( t \in [0, 1] \) is considered a random process and is assumed to be integrable over the unit interval \([0, 1]\).

Suppose we know the true number of clusters, then we extend the traditional measurement error model [\cite{fuller2009measurement}] to the clustered case. The functional data are assumed to come from $C$ different clusters. In particular, for $i$th $(i=1,2,...n_c)$ subject and $j$th  $(j=1,...,J)$ repeated measure in the $c$th $(c=1,...,C)$ cluster, we assume
\begin{equation}
W_{cij}(t) = X_{ci}(t) + U_{cij}(t).    
\end{equation}
Here, the true functional measurement \( X_{ci}(t) \) are unobserved, and \( W_{cij}(t) \) represents the \( j \)th replicate of observed measurements for subject \( i \) in the \( c\)th cluster, and is contaminated by an unknown functional measurement error \( U_{cij}(t) \). The observed values\( W_{cij}(t) \) is an unbiased surrogate for the true value \( X_{ci}(t) \). The true values \( X_{ci}(t) \) and measurement errors \( U_{cij}(t) \) are uncorrelated.

\xl{We assume that subjects within the same cluster share a common mean trajectory, denoted by \(\mu_c(t)\). For each subject \(i = 1, \ldots, n_c\) in cluster \(c\), the observed trajectory can be expressed as  
\[
X_{ci}(t) = \mu_c(t) + b_{ci}(t),
\]  
where \(b_{ci}(t)\) is a random term representing the individual's deviation from the mean trajectory.}

Consequently, for any fixed $t\in[0,1]$, the pointwise clustered mixed effect model with measurement error has the form  
\begin{equation}
W_{cij}(t) = 
   \alpha_{c}(t)+b_{ci}(t)+U_{cij}(t),
\end{equation}
where $\alpha_{c}(t)$ is fixed intercept for the cluster c, $b_{ci}(t)$ is random intercept for subject i within cluster c. In addition, we assume  $b_{ci}(t) \sim N(0,\sigma_{c}^2)$ and $\epsilon_{cij}(t)\sim N(0,\sigma^2)$. For any fixed $t$, $\left\{b_{ci}(t)\right\}$ and $\left\{U_{cij}(t)\right\}$ are all independent of each other. 

\xl{When the clustering index is completely known, a point-wise linear mixed effect model can be used to obtain the predicted values for the latent $X_{ci}(t)$} based on the repeated measurements $W_{cij}(t)$. This approach yields the predicted trajectory for \( \hat{X}(t) \) at each point $t$. If the true number of clusters is unknown, clustered results using single measures or averaged data from J replicated measures, or prior clustered information could be used for initial cluster groups.

The high or infinite dimensionality of the estimated functional matrices poses challenges for estimation. To address this, polynomial spline basis expansion is applied to reduce dimensionality, \xl{also with smoothing techniques incorporated to account for the temporal correlation structure in \( \hat{X}(t) \) [\cite{reinsch1967smoothing}]. } 
Temporal correlation is modeled through smoothing penalties, which discourage large differences between adjacent spline coefficients, thereby enforcing continuity and gradual transitions in the estimated curve \(\hat{X}(t)\). The placement and number of basis functions (\(K_n\)) to balance the model’s flexibility and smoothness, ensuring that the estimate captures meaningful temporal patterns without overfitting noise[\cite{wang1998smoothing,krivobokova2007note}].

\begin{equation*}
\hat{X}(t) = 
\Sigma_{k=1}^{K_n}\gamma_kb_k(t),
\end{equation*}
where $\{\gamma_k\}_{k=1}^{K_n}$ are unknown spline coefficients and $\{b_k(t)\}_{k=1}^{K_n}$ are a series of spline basis functions defined on the unit interval. The number of bases, $K_n$, is selected based on the sample size n. $\{\gamma_k\}_{k=1}^{K_n}$ are used for next clustering step.

\subsection{Clustering}

The spline coefficients $\{\gamma_k\}_{k=1}^{K_n}$ are used for final clustering. In this paper, a Gaussian Mixture Modeling for Model-Based Clustering (MCLUST) method is used for clustering the optimal number of clusters, and each observation is assigned to one of the C clusters[\cite{fraley1998many}]. The method relies on the Expectation-Maximization (EM) algorithm to iteratively estimate the parameters of each cluster, aiming for maximum likelihood. MCLUST uses statistical criteria, like the Bayesian Information Criterion (BIC), to determine the number of clusters. Other clustering methods, based on different data distributions, can also be employed.

**Model Assumption**: Each observation \( \mathbf{x}_i \) is generated from one of \( C \) components in a Gaussian mixture model for $i$th $(i=1,2,...n_c)$ subject,
   \begin{equation*}
   f(\mathbf{\gamma}_i \mid \theta_c) = \sum_{c=1}^C \pi_c \cdot \mathcal{N}(\mathbf{\gamma}_i \mid \mu_c, \Sigma_c)
   \end{equation*}
where \( \pi_c \) is the mixture proportion, \( \mu_c \) is the mean, and \( \Sigma_c \) is the covariance matrix for each component \( c \).

**Covariance Structure**: Covariance matrices \( \Sigma_c \) are parametrized as:
   \begin{equation}
    \Sigma_c = \lambda_c D_c A_c D_c^T
   \end{equation}
 where \( D_c \) represents orientation, \( A_c \) represents shape, and \( \lambda_c \) the volume of clusters. This enables flexibility in modeling various cluster shapes.

**Expectation-Maximization (EM) Algorithm**: To maximize the likelihood:

   - E-step: Calculate the posterior probability \( \tau_{ic} \) that observation \( i \) belongs to component \( c \):

   \begin{equation}
   \tau_{ic} = \frac{\pi_c f(\mathbf{x}_i \mid \mu_c, \Sigma_c)}{\sum_{j=1}^C \pi_j f(\mathbf{x}_i \mid \mu_j, \Sigma_j)}    
   \end{equation}
   
   - M-step: Update \( \mu_c \), \( \Sigma_c \), and \( \pi_c \) to maximize the expected complete-data log-likelihood  \( Q(\Theta \mid \Theta^{(t)}) \), given the observed data and the current parameter estimates \( \Theta^{(t)} \).

   \begin{equation}
     Q(\Theta \mid \Theta^{(t)}) = \sum_{i=1}^n \sum_{c=1}^C \tau_{ic} \left( \log \pi_c + \log \mathcal{N}(\mathbf{x}_i \mid \mu_c, \Sigma_c) \right)  
   \end{equation}
where the term \( \tau_{ic} \left( \log \pi_c + \log \mathcal{N}(\mathbf{x}_i \mid \mu_c, \Sigma_c) \right) \) represents the contribution of each observation to the log-likelihood, weighted by its probability of belonging to component \( c \). \(\Theta^{(t)}\) is the parameter estimate after the 
t-th iteration.

**Model Selection**: Bayesian Information Criterion (BIC) is used to select the best model and number of clusters:
   
   \begin{equation}
   BIC = 2 \log L - m \log(n)       
   \end{equation}

   where \( L \) is the maximized likelihood, \( m \) is the number of parameters, and \( n \) the sample size.

\subsection{Algorithm}
The whole clustering procedure proceeds as follows.

\begin{longlist}
\item[1.] Obtain an initial clustering result based on either a single measure \( W_{ij}(t) \) or averaged values of J replicates measures.
\item[2.] Apply the CMEM-ME method to estimate predicted values \( \hat{X}(t) \).
\item[3.] Incorporate the temporal correlation structure in \( \hat{X}(t) \) with smoothing techniques, here we use basis expension and get coefficient \(\{\gamma_k\}_{k=1}^{K_n}\) .
\item[4.] Perform cluster analysis on \(\{\gamma_k\}_{k=1}^{K_n}\).
\item[5.] Repeat step 2-4 until

the subject memberships remain unchanged between the \(r\)th iteration and the \((r - 1)\)th iteration.
\end{longlist}

\section{Simulation}
\subsection{Simulation procedure}

Simulation studies were used to evaluate the finite sample performance of the proposed methods. The true functional values were generated from three groups with
\[
X_{i}(t) = \begin{cases}
\sin(2\pi t)+\epsilon_i(t),\; \text{ for } \; i=1,...,n/3 \\
\frac{\sin\left(\pi \left(16 \left(t - 0.5\right)\right) / 2\right)}{1 + \left(2 \left(16 \left(t - 0.5\right)\right)^2\right) \left(\sin\left(8 \left(t - 0.5\right)\right) + 1\right)}+\epsilon_i(t),\; \text{ for } \; i=n/3+1,...,2n/3 \\
1 - 2\exp(-6t)+\epsilon_i(t),\; \text{ for } \; i=2n/3+1,...,n
\\
\end{cases}
\]
The error term \(\epsilon_i(t)\) was simulated independently from a Gaussian process with a mean of zero, constant variance \(\sigma_X^2\), and covariance function \(\Sigma_X(s,t) = \sigma_X^2 \exp\left(-\frac{(s-t)^2}{2\rho_X^2}\right)\), where the correlation depends on the distance between two points and \(\rho_X\) controls the strength of correlation. The observed functional values \(W_{cij}(t)\) were simulated using the traditional additive measurement error model: \(W_{ij}(t) = X_{i}(t) + U_{ij}(t)\), where \(U_{ij}(t)\) was simulated independently from a Gaussian process with a mean of zero, constant variance \(\sigma_U^2\), and covariance function \(\Sigma_U(s,t) = \sigma_U^2 \exp\left(-\frac{(s-t)^2}{2\rho_U^2}\right)\), where the correlation depends on the distance between two points and $\rho_U$ controls the strength of correlation. In this simulation, five replicated measurements were generated for each individual.

The simulation studies were conducted to evaluate the performance of the proposed estimators under the following conditions (1) increasing sample sizes with n = 300, 600, 1500 or 3000; (2) different magnitudes of measurement error with \(\sigma_X= 0.5\) and \(\sigma_U= 0.5,1,1.5,2\) or \(\sigma_X= 1\) and \(\sigma_U= 1,1.5,2,3\) or \(\sigma_X= 1.5\) and \(\sigma_U= 1.5,2,2.5,3\); (3) different strengths of correlation with \(\rho_X=0.25,0.5,0.75\) and \(\rho_U=0.25,0.5,0.75\); (4) different structures for the variance-covariance matrix of the true functional values and measurement error: including compound symmetric, autoregressive one and Independent. In our simulation, the number of replications is 500.

We consider four estimation methods, namely, the oracle, the naive, the average, and the functional mixed-effects model with clustering(FMEC). The oracle estimator was obtained based on the true functional values \(X_i(t)\). The naive estimator was obtained by using one of the observed measurements among five repeated measures. In the simulation, we use the first replicate \(W_i1(t)\) for the Naive estimator.  The average estimator was obtained by averaging all five replicates of observed measurements as \(\bar{W_{ij}}(t)\). 

\subsection{Index}
We use the following three indexes to compare the numerical performances of the different clustering methods.

\textbf{Average Group Size}. In simulation and practical studies, we use MCLUST package in R to perform the clustering. Here, MCLUST utilizes functions that integrate Expectation-Maximization (EM) and the Bayesian Information Criterion (BIC) to automatically determine the optimal number of clusters [\cite{fraley1998mclust}]. In our simulation, this average number of clusters \(S\) over $500$ replications is calculated as
\[Asize = \frac{1}{500}\sum_{r=1}^{500} S_r\]

\textbf{Rand index, adjusted Rand Index and Jaccard index}. As the true membership of samples were known, we could assess the performance of estimation methods externally by computing the Rand index [\cite{rand1971objective}], adjusted Rand index [\cite{hubert1985comparing}] and Jaccard index [\cite{jaccard1912distribution}]. These three indexes are used to evaluate clustering performance because they either adjust for chance agreement or focus on true positive pairs, suitable for imbalanced datasets [\cite{pinto2007ranked,romano2016adjusting,gupta2015significance}]. For a given clustering membership of subjects, the concepts of true positive (TP), true negative (TN), false positive (FP), and false negative (FN) are defined as follows. TP refers to the number of subject pairs that are correctly identified as belonging to the same cluster. TN denotes the number of subject pairs that are correctly identified as belonging to different clusters, meaning they originate from separate clusters and are assigned to distinct clusters. FP represents the number of subject pairs that originate from different clusters but are incorrectly assigned to the same cluster. FN indicates the number of subject pairs that belong to the same cluster but are incorrectly assigned to different clusters. Therefore, the Rand index 
\(Rand=\frac{TP+TN}{TP+FN+FP+TN}\) and the adjusted Rand Index 
\(aRand = \frac{Rand - E(Rand)}{max(Rand)-E(Rand)}\). The Jaccard index \(Jaccard=\frac{TP}{TP+FN+FP}\) .
For all three indexes, higher values signify better agreement between the estimated and true group memberships. 

\subsection{Simulation result}
\subsubsection{Effects of varying sample sizes}

\begin{figure}[H]
    \centering
    \includegraphics[width=\linewidth]{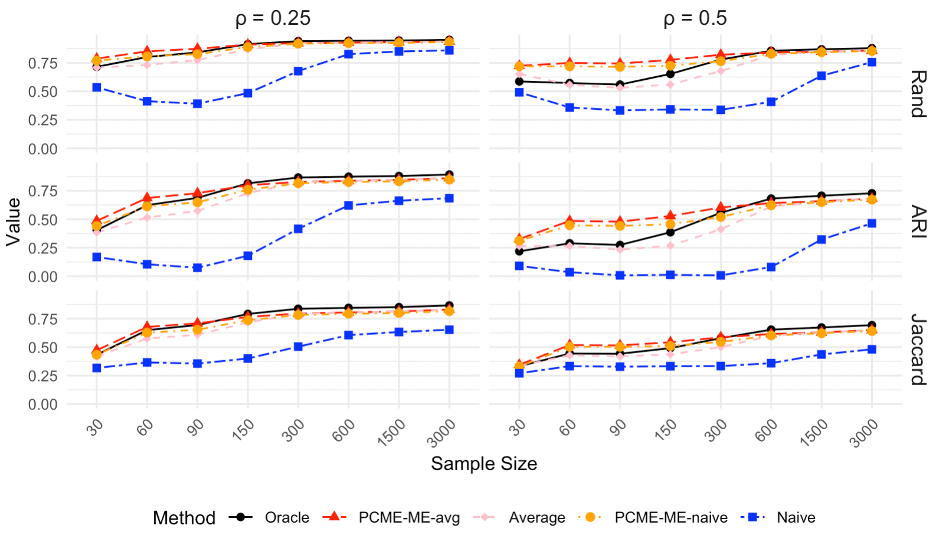}
    \caption{The effect of varying sample sizes on the clustering performance with the Oracle, PCME-ME-avg, Average, PCME-ME-naive and Naive estimators is shown by Rand Index, Adjusted Rand Index (ARI) and average Jaccard Index for \( n \in \{30, 60, 90, 150, 300, 600, 1500, 3000\} \). Oracle is benchmark, PCME-ME-avg is point clustered mixed effect model with averaged clusters as initial value, Average is averaging all replicates of data, PCME-ME-naive is point clustered mixed effect model with naive clusters as initial value and Naive is one replicate of data. The parameters are set as \(\sigma_X = 1.5\), \(\sigma_U = 1.5\), and \(\rho \in \{0.25, 0.5\}\), where \(\rho\) is the correlation parameter in the AR(1) covariance function of the functional true values and measurement error. }
    \label{compare_samplesize}
\end{figure}

We provide the simulation results for evaluating the impacts of varying sample sizes on the clustering performance with the Oracle, PCME-ME-avg, Average, PCME-ME-naive, and Naive methods in the Figure ~\ref{compare_samplesize}. Under \(\rho = 0.25\), where the correlation in the AR(1) covariance structure is weaker, the clustering performance of all methods improves as the sample size increases. For larger sample sizes (\(n \geq 150\)), the benchmark Oracle technique consistently yields the top values for all metrics (Rand Index, ARI, and Jaccard Index). It's interesting to see that PCME-ME-avg outperforms Oracle in small sample sizes (\(n \leq 150\)). PCME-ME-avg gets closer to Oracle as the sample size grows, especially for bigger datasets (\(n \geq 600\)). Although PCME-ME-naive performs well as well, it lags behind PCME-ME-avg by a small margin, suggesting that the capacity to recover real clusters is influenced by the initial clustering values chosen. For small to moderate sample sizes (\(n \leq 300\), the Average technique performs rather well, surpassing the Naive approach but falling short of the PCME-ME approaches' accuracy no matter for what initial values, indicating the essential need to use our method. Even with increasing sample size, the Naive method, performs the worst and exhibits significant clustering error, demonstrating its inability to handle measurement error effectively.

Under \(\rho = 0.5\), where the correlation between measurements is stronger, all methods based on higher temporal structure, leading to weeker clustering performance across metrics.  PCME-ME-avg remains the most robust among the error-prone methods, achieving performance comparable to Oracle for larger sample sizes (\(n \geq 600\)) and outperforming Oracle for smaller and moderate sample sizes. PCME-ME-naive also shows substantial improvement with increasing sample size, though it slightly underperforms PCME-ME-avg, it still outperms than Oracle and Average method when (\(n \leq 300\). Overall, the Average technique is less successful than the PCME-ME methods, especially for small and moderate sample sizes (\(n = 60 \) to \(150\)). Even though the larger correlation helps the Naive technique a little, it still performs badly for all sample sizes, highlighting its shortcomings in dealing with measurement error.

Both \(\rho = 0.25\) and \(\rho = 0.5\) show how crucial sample size is in reducing the performance difference among approaches. The PCME-ME techniques (avg and naive initial values) show a discernible improvement with small sample sizes, especially in situations with stronger correlation. When sample size is large enough, it doesn't matter which method to use for correcting measurement error.

\subsubsection{The effect of measurement error}

\begin{figure}[H]
    \centering
    \includegraphics[width=\linewidth]{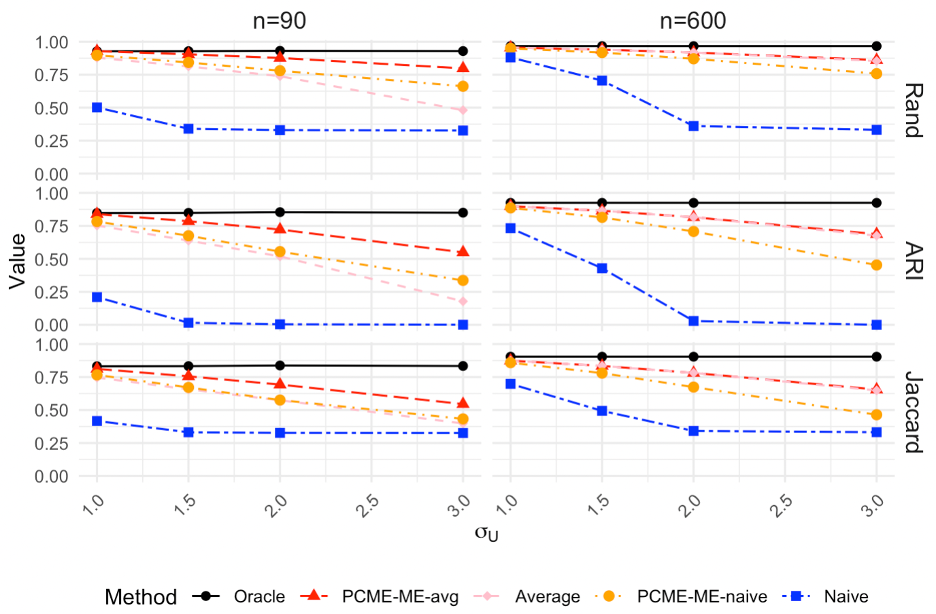}
    \caption{The effect of varying measurement error on the clustering performance with the Oracle, PCME-ME-avg, Average, PCME-ME-naive and Naive estimators is shown by Rand Index, Adjusted Rand Index (ARI) and average Jaccard Index for \( \sigma_U \in \{1, 1.5, 2, 3\} \). Oracle is benchmark, PCME-ME-avg is point clustered mixed effect model with averaged clusters as initial value, Average is averaging all replicates of data, PCME-ME-naive is point clustered mixed effect model with naive clusters as initial value and Naive is one replicate of data. The parameters are set as \(\sigma_X=1 \), \(n \in \{90, 600\}\), and \(\rho= 0.5\), where \(\rho\) is the correlation parameter in the AR(1) covariance function of the functional true values and measurement error.}
    \label{compare_error}
\end{figure}

In order to analyze the outcomes for assessing how different measurement errors affect the clustering performance using the Oracle, PCME-ME-avg, Average, PCME-ME-naive, and Naive estimators, we run simulations among \( n = 1500 \), \( \sigma_X = 1.5 \), \( \rho \in \{0.25, 0.5\} \), and \( \sigma_U \in \{1.5, 2, 2.5, 3\} \) in Figure ~\ref{compare_error}.
The performance of all approaches declines as \( \sigma_U \) rises for lower sample sizes (\( n = 90 \)), underscoring the detrimental effects of increased measurement error. As evidenced by no additive measurement error, the Oracle approach continuously produces the best results across all metrics (Rand Index, ARI, and Jaccard Index), and the values remain the same. Among the error-prone methods, PCME-ME-avg exhibits the best performance, closely approximating the Oracle for lower values of \( \sigma_U \) and maintaining relatively higher values for all metrics as \( \sigma_U \) increases. Although it lags slightly behind PCME-ME-avg, PCME-ME-naive likewise performs well, especially with larger measurement error, suggesting better way than averaging values. With notable drops in clustering accuracy across all measures as \( \sigma_U \) rises, the Naive method performs the worst, highlighting its shortcomings in mitigating measurement error in smaller datasets.

For larger sample sizes (\( n = 600 \)), all methods exhibit improved performance compared to \( n = 90 \), with smaller declines in clustering accuracy as \( \sigma_U \) increases. The Oracle approach achieves nearly flawless values across all criteria. Even at higher \( \sigma_U \) levels, PCME-ME-avg maintains its superior performance over other error-prone techniques, demonstrating resistance to measurement error. Its scalability and dependability in larger datasets are demonstrated by the near alignment of its performance with Oracle, especially for metrics like ARI and Jaccard Index. The Average method performs significantly better, even outperforming the PCME-ME-naive method in larger datasets. The Naive approach continues to perform the worst, demonstrating its incapacity to control measurement error even in larger datasets. According to scenarios with varying sample sizes, PCME-ME-avg may largely correct measurement error and get results that are close to Oracle.

\subsubsection{The effect of correlation}

\begin{figure}[H]
    \centering
    \includegraphics[width=\linewidth]{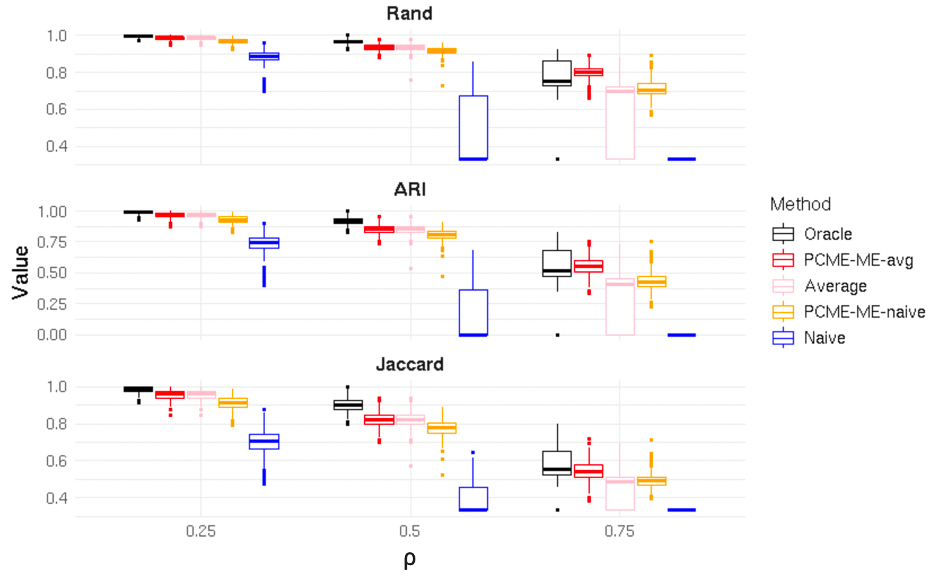}
    \caption{The effect of varying correlation on the clustering performance with the Oracle, PCME-ME-avg, Average, PCME-ME-naive and Naive estimators is shown by Rand Index, and Average Rand Index (ARI) and Jaccard Index  for \( n \in \{300\} \). Oracle is benchmark, PCME-ME-avg is point clustered mixed effect model with averaged clusters as initial value, Average is averaging all replicates of data, PCME-ME-naive is point clustered mixed effect model with naive clusters as initial value and Naive is one replicate of data. The parameters are set as \(\sigma_X = 1\), \(\sigma_U = 1.5\), and \(\rho \in \{0.25, 0.5, 0.75\}\), where \(\rho\) is the correlation parameter in the AR(1) covariance function of the functional true values and measurement error. }
    \label{compare_correlation}
\end{figure}

Simulation findings on the effects of changing the correlation parameter (\(\rho\)) on the clustering performance of Oracle, PCME-ME-avg, Average, PCME-ME-naive and Naive estimators are shown in Figure~\ref{compare_correlation}. The correlation parameter in the AR(1) covariance function of the functional true values and measurement error is represented by \(\rho\). Each boxplot represents the distribution of values from 500 iterations conducted under the simulation scenario with \(\sigma_X = 1\), \(\sigma_U = 1.5\) and \( n = 300 \). An rise in the correlation parameter (\(\rho\)) in the AR(1) configuration, particularly from the Jaccard Index, was associated with a considerable decline in clustering performance. The PCME-ME-avg and Average methods perform similarly, approaching the Oracle technique, in smaller correlations. When \(\rho\) rose to 0.75, it was evident that both PCME-ME methods had excellent clustering accuracy and low variability; in fact, PCME-ME-avg even had greater median values and smaller interquartile ranges than Oracle approach. The Average Method has significantly greater interquartile ranges and is situated between the PCME-ME and Naive estimators. This consistent performance shows how well PCME-ME corrects measurement errors as the correlation parameter increases.

\subsubsection{The effect of functional error term} 

\begin{figure}[H]
    \centering
    \includegraphics[width=\linewidth]{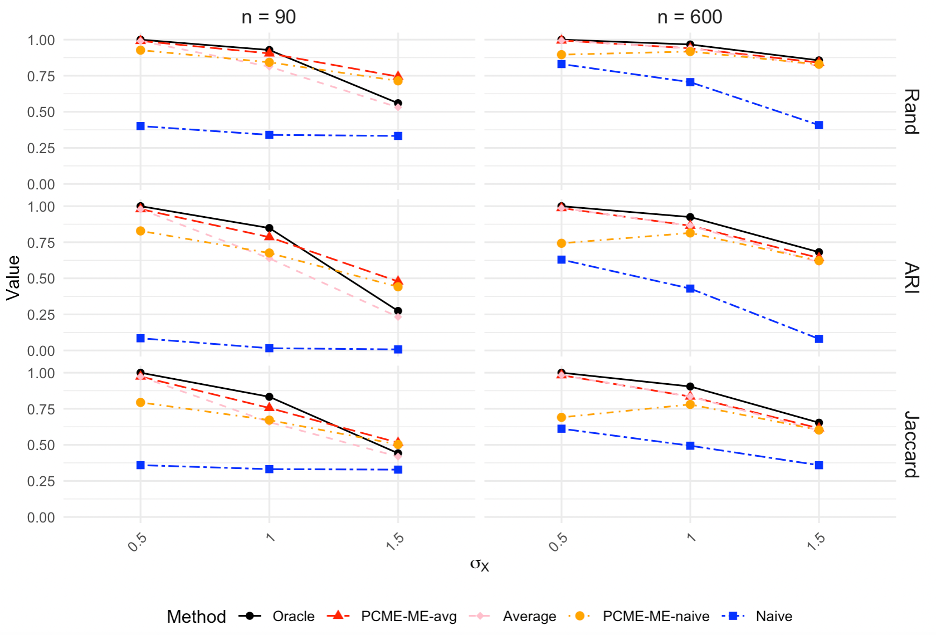}
    \caption{The effect of varying functional error on the clustering performance with the Oracle, PCME-ME-avg, Average, PCME-ME-naive and Naive estimators is shown by Rand Index, Adjusted Rand Index (ARI) and average Jaccard Index for \( \sigma_X \in \{0.5, 1, 1.5\} \). Oracle is benchmark, PCME-ME-avg is point clustered mixed effect model with averaged clusters as initial value, Average is averaging all replicates of data, PCME-ME-naive is point clustered mixed effect model with naive clusters as initial value and Naive is one replicate of data. The parameters are set as \(\sigma_U=1.5 \), \(n \in \{90, 600\}\), and \(\rho= 0.5\), where \(\rho\) is the correlation parameter in the AR(1) covariance function of the functional true values and measurement error.}
    \label{compare_functional}
\end{figure}

To provide the results for evaluating the impacts of varying the functional error term (\(\sigma_X\)) on the clustering performance with the Oracle, PCME-ME-avg, Average, PCME-ME-naive and Naive estimators, we made simulation in \( n = 90, 600 \), \(\sigma_U = 1.5\), and \(\rho = 0.5\), where \(\rho\) is the correlation parameter in the AR(1) covariance function of the functional true values and measurement error. The figure shows in Figure~\ref{compare_functional}. When the functional error term (\(\sigma_X\)) increased, the performance metrics of all estimators consistently decreased in both lower and bigger sample sizes. Clustering performance will deteriorate as the dispersion within a cluster increases, yet PCME-ME-avg consistently manages this disparity well. When the difference within the same cluster is minor (\(\sigma_X = 0.5\)), it is desirable to average all replicate data to adjust for measurement errors. The Naive method performs the worst overall, particularly when the sample size is small, and the clustering performance is poor regardless of the extent of the functional error.

\subsubsection{The effect of variance covariance matrix}

Simulation findings of clustering performance across various variance-covariance structures, including AR(1), Independent, and Compound Symmetry, are shown in Table~\ref{covariance}. \( n = 90 \), \(\sigma_X = 1\), \(\sigma_U \in \{1, 1.5, 2, 3\}\), and \(\rho = 0.5\) are the parameters that are set. For the AR(1) covariance structure**, the Oracle method consistently achieves the highest values across all three performance metrics, and the PCME-ME-avg method closely follows the Oracle method in performance, performing better than the PCME-ME-naive, Average, and Naive methods. In the Independent covariance structure, the clustering methods achieve higher overall performance compared to AR(1), and the performance rate doesn't change a lot when measurement error increases. The PCME-ME-avg and Average methods perform similarly, and the PCME-ME-naive  method performs much better than Naive method. When initial clustering value is undefined, Using PCME-ME method could always improve the clustering performance. For the Compound Symmetry covariance structure, as the measurement error increases, the Average method outperforms other methods. However, when we examine larger sample sizes, particularly when \( n \geq 300 \), the performance of PCME-ME-avg and Average methods becomes comparable. The results highlight how crucial it is to use suitable clustering techniques that accurately account for measurement error in terms of correlation structures in order to guarantee reliable clustering of functional data.

\begin{table}[H]
  \centering
  \caption{Clustering performance with different variance-covariance structures: AR(1), Independent, and Compound Symmetry. The performance metrics are shown by the average Rand Index (Rand), average Adjusted Rand Index (ARI), and average Jaccard Index (Jaccard). Oracle is the benchmark, PCME-ME-avg is the point clustered mixed effect model with averaged clusters as the initial value, Average is the averaging of all replicates of data, PCME-ME-naive is the point clustered mixed effect model with naive clusters as the initial value, and Naive is one replicate of data. The parameters are set as $n = 90$, $\sigma_X = 1.5$, and $\rho = 0.5$.}
  \resizebox{\textwidth}{!}{
  \begin{tabular}{r|rrrrr|rrrrr|rrrrr}
    \toprule
          & \multicolumn{5}{c|}{Rand} & \multicolumn{5}{c|}{ARI} & \multicolumn{5}{c}{Jaccard} \\
    \midrule
    $\sigma_U$ & Oracle & PCME-ME-avg & Average & PCME-ME-naive & Naive & Oracle & PCME-ME-avg & Average & PCME-ME-naive & Naive & Oracle & PCME-ME-avg & Average & PCME-ME-naive & Naive \\
    \midrule
    \multicolumn{16}{c}{\textbf{AR(1)}} \\
    1.0 & 0.928 & 0.928 & 0.879 & 0.897 & 0.502 & 0.848 & 0.840 & 0.753 & 0.783 & 0.210 & 0.832 & 0.812 & 0.746 & 0.766 & 0.417 \\
    1.5 & 0.928 & 0.905 & 0.813 & 0.843 & 0.340 & 0.849 & 0.785 & 0.638 & 0.675 & 0.016 & 0.833 & 0.756 & 0.657 & 0.672 & 0.332 \\
    2.0 & 0.931 & 0.877 & 0.738 & 0.780 & 0.330 & 0.854 & 0.722 & 0.519 & 0.555 & 0.004 & 0.838 & 0.693 & 0.574 & 0.576 & 0.327 \\
    3.0 & 0.929 & 0.798 & 0.481 & 0.662 & 0.327 & 0.850 & 0.549 & 0.178 & 0.337 & 0.001 & 0.834 & 0.545 & 0.398 & 0.433 & 0.326 \\
    \midrule
    \multicolumn{16}{c}{\textbf{Independent}} \\
    1.0 & 0.999 & 0.998 & 0.998 & 0.997 & 0.983 & 0.998 & 0.996 & 0.996 & 0.994 & 0.962 & 0.998 & 0.995 & 0.995 & 0.992 & 0.952 \\
    1.5 & 0.999 & 0.997 & 0.997 & 0.989 & 0.891 & 0.998 & 0.992 & 0.992 & 0.975 & 0.778 & 0.998 & 0.990 & 0.990 & 0.967 & 0.770 \\
    2.0 & 0.999 & 0.993 & 0.993 & 0.966 & 0.600 & 0.998 & 0.984 & 0.984 & 0.923 & 0.337 & 0.998 & 0.979 & 0.979 & 0.903 & 0.482 \\
    3.0 & 0.999 & 0.975 & 0.972 & 0.887 & 0.342 & 0.998 & 0.943 & 0.938 & 0.742 & 0.017 & 0.998 & 0.928 & 0.923 & 0.712 & 0.332 \\
    \midrule
    \multicolumn{16}{c}{\textbf{Compound Symmetry}} \\
    1.0 & 0.992 & 0.991 & 0.990 & 0.992 & 0.963 & 0.981 & 0.978 & 0.977 & 0.982 & 0.915 & 0.976 & 0.972 & 0.971 & 0.977 & 0.898 \\
    1.5 & 0.992 & 0.983 & 0.987 & 0.986 & 0.798 & 0.981 & 0.960 & 0.971 & 0.967 & 0.639 & 0.976 & 0.952 & 0.963 & 0.958 & 0.678 \\
    2.0 & 0.992 & 0.966 & 0.981 & 0.967 & 0.484 & 0.981 & 0.918 & 0.955 & 0.922 & 0.200 & 0.976 & 0.907 & 0.944 & 0.904 & 0.423 \\
    3.0 & 0.992 & 0.904 & 0.948 & 0.888 & 0.335 & 0.982 & 0.784 & 0.884 & 0.738 & 0.010 & 0.976 & 0.785 & 0.867 & 0.712 & 0.329 \\
    \bottomrule
  \end{tabular}%
  }
  \label{covariance}%
\end{table}%

\section{Data application}
\subsection{Physical activity intensity of NHANES data}

We applied data from the National Health and Nutrition Examination Survey (NHANES), a survey that provides a wealth of health and nutrition data, to apply the suggested FMEC model. Two survey cycles (2003–2004 and 2005–2006) are used in this research, and data on physical activity was gathered using the Physical Activity Monitor (PAM)\cite{CDC_NHANES}. Activity intensity was captured by the PAM device at 1-minute intervals, and the intensity data was then aggregated by hour. The dataset also contained other demographic factors.

The inclusion criteria for this analysis were as follows:
\begin{longlist}
    \item[1] Participants aged 20 years or older,
    \item[2] Participants with complete 7-day, 24-hour physical activity data, and
    \item[3] Participants with no missing values for relevant demographic variables.
\end{longlist}

A total sample of 2,196 people was kept after these criteria. Because physical activity levels are usually lower between 10:00 pm and 6:00 am than at other times, data from this time frame were excluded for analysis. 

\begin{figure}[h]
    \centering
    \includegraphics[width=\linewidth]{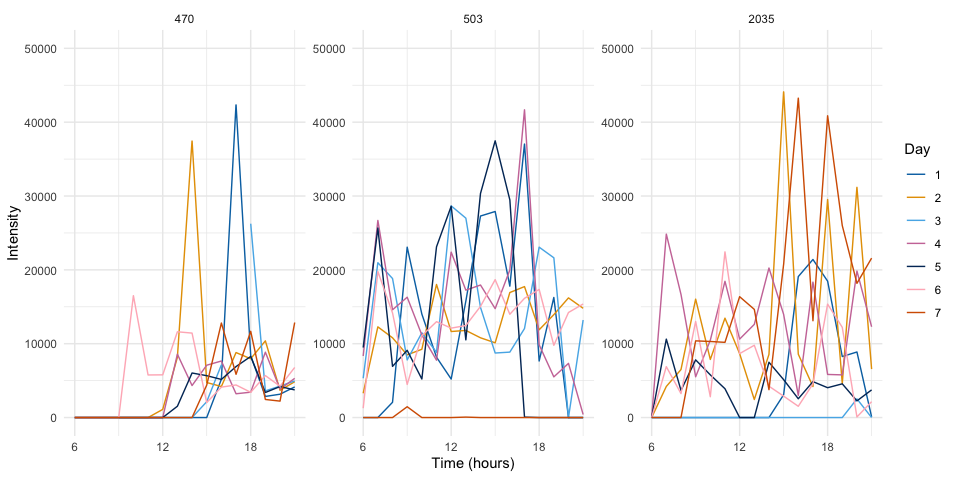}
    \caption{Grouped intensity trajectories among different methods}
    \label{sample_NHANES}
\end{figure}

The physical activity intensity trajectories for three randomly chosen subjects (470, 503, and 2035) throughout a seven-day period, measured over a 16-hour period from 6 AM to 10 PM, are shown in Figure ~\ref{sample_NHANES}. Every participant displays unique patterns of activity, with significant variation in intensity within and between subjects. Subject 2035, for example, has consistently greater activity levels with noticeable spikes on Days 2 and 7, whereas Subject 470 exhibits generally very moderate intensity with sporadic peaks on particular days. Subject 503 has a more similar pattern, with daily variations in intensity. The daily variations are highlighted by the intensity trajectories within each individual. The intensity value also vary greatly; some readings are above 40,000 units, while others stay close to zero all day. Further investigation into clustering, variability, and behavioral tendencies in physical activity data, as well as possible measurement inaccuracy in wearable devices, is made possible by this plot, showing the variability in physical activity patterns between people and days.

Three methods—the FMEC method, the Average method, and the Naive method—were then employed to correct for measurement error in physical activity intensity data. For clustering, the \texttt{Mclust} package in R was utilized, categorizing participants into intensity groups ordered from lowest to highest. These intensity group memberships were subsequently used as predictor variables to examine their relationship with diabetes status, controlling for gender, ethnicity, and age group. For the naive approach, days 2 and 7 were chosen as representative random days. The clustering results are shown in Table~\ref{group1}. 
Different days would produce different number of clusters. While FMEC method get 6 groups, Average method clusters for 9 groups, while Naive method only obtains 3 or 4 groups.

\begin{table}[H]
  \centering
  \caption{Group tables among different methods}
    \begin{tabular}{lrrrrrrrrr}
    \toprule
    Group & 1     & 2     & 3     & 4     & 5     & 6     & 7     & 8     & 9 \\
    \midrule
    FMEC  & 73    & 220   & 412   & 269   & 522   & 700   &       &       &  \\
    Average   & 66    & 69    & 238   & 274   & 276   & 345   & 408   & 263   & 257 \\
    Naive (D2) & 153   & 253   & 806   & 984   &       &       &       &       &  \\
    Naive (D7) & 188   & 980   & 1028  &       &       &       &       &       &  \\
    \bottomrule
    \end{tabular}
  \label{group1}
\end{table}

\begin{figure}[H]
    \centering
    \includegraphics[width=\linewidth]{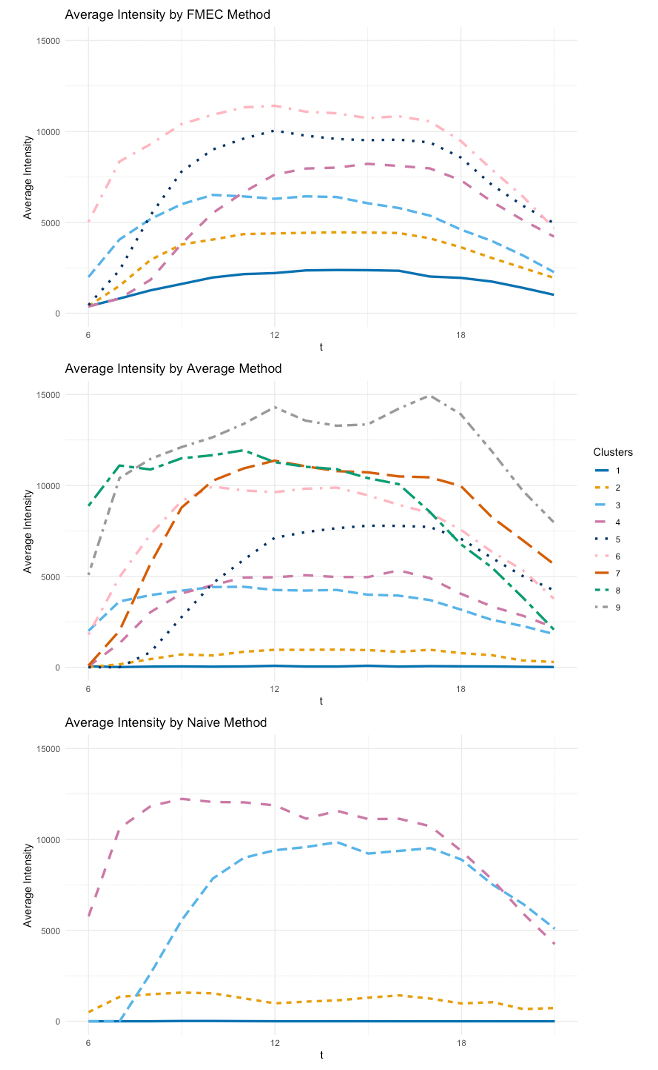}
    \caption{Grouped intensity trajectories among different methods}
    \label{trajectory_NHANES}
\end{figure}

Figure ~\ref{trajectory_NHANES} illustrates the average physical activity intensity throughout the days (from 6 am to 10 pm) across clusters determined by three methods: the FMEC method, the Average method and the Naive method. Each plot reveals distinct patterns of activity peaking around midday and gradually decreasing in the evening. 
In the first subplot, the FMEC approach divides the data into six clusters. Without unduly complicating the analysis, this strategy better captures the diversity in activity levels while maintaining the general circadian regularity observed in the later methods. For example, clusters 1 and 2 keep the lowest intensities, while cluster 6 represents the greatest intensity group, similar with cluster 9 in the Average approach. FMEC offers a more subtle clustering than the Naive approach and is easier to understand than the average method, which has fewer clusters. 
The second subplot provides a more detailed perspective of activity patterns by grouping the data into nine groups using the Average method. Particularly for intermediate activity groups like clusters 3 to 6, the extra clusters capture minute variations in intensity levels. Compared to the Naive method, the trajectories are more precise, but the greater number of clusters can make interpretation more difficult. 
The third subplot displays a general rhythm in physical activity by using the Naive approach to group the data into four clusters. In clusters 3 and 4, intensity levels rise in the morning (after 6:00), peak at noon (12:00), and then level out in the evening (after 18:00). Cluster 1 continuously maintains lowest throughout the day, whereas cluster 4 constantly displays the highest levels. The simplicity of the basic approach, however, leads to fewer clusters and may obscure subtler differences in activity patterns. All things considered, the FMEC approach seems to provide the best trade-off between interpretability and complexity.

Table ~\ref{prop_nhanes} compares the distribution of demographic variables across clusters derived using three different clustering methods: FMEC method, Avereage Method and Naive Method. The FMEC method, compared to Average and Naive method, offers more distinct and meaningful clusters. For example, cluster 1 from the FMEC method is characterized by a significantly higher percentage of individuals aged $\le$ 44 (68.5\%) compared to other clusters. Moreover, the FMEC method shows clearer differences in ethnicity composition between clusters, when cluster 3 having a higher proportion of White individuals (80.8\%) and cluster 4 showing a more diverse ethnic composition. Gender patterns in FMEC method are also more distinct, as shown by cluster 6, which has the largest percentage of males (35.3\%). With nine clusters, the Average method offers more detail in the distributions of age and ethnicity. Nonetheless, certain clusters, including clusters 3, 4, and 5, which share mean ages and age distributions. Since the clusters for the Naive method are loosely categorized, subtle differences are not captured. The ethnic distribution under the Naive method, for example, is comparatively homogeneous, with "White" being the predominant group in every cluster. And there are no notable variations in the age distribution amongst clusters, and the mean age values fall between 42.3 and 56.8.

\begin{table}[H]
  \centering
  \tiny
  \caption{Proportion tables for different clustering methods}

  \vspace{0.5em}
  \textbf{Proportion table for FMEC method}
  \vspace{0.5em}
  \begin{tabular}{ccccccc}
    \textbf{Intensity} & \textbf{Count} & \textbf{Diabetes} & \textbf{Gender} & \textbf{Ethnicity} & \textbf{Age Group} & \textbf{Age} \\
    group  &       & No | Yes & Male | Female & White | Hispanic | Black | Others & $\leq 44$ | 45-64 | $\geq 65$ &     \\
    1     & 73    & 94.5 | 5.5 & 45.2 | 54.8 & 74 | 13.7 | 8.2 | 4.1 & 68.5 | 21.9 | 9.6 & 39.6 \\
    2     & 220   & 77.7 | 22.3 & 42.7 | 57.3 & 77.3 | 7.3 | 12.7 | 2.7 & 31.4 | 16.8 | 51.8 & 60.2 \\
    3     & 412   & 80.6 | 19.4 & 46.4 | 53.6 & 80.8 | 6.1 | 10.2 | 2.9 & 19.2 | 17.7 | 63.1 & 64.7 \\
    4     & 269   & 84 | 16 & 32.3 | 67.7 & 68.8 | 11.5 | 15.2 | 4.5 & 37.5 | 27.5 | 34.9 & 52.1 \\
    5     & 522   & 87 | 13 & 32.8 | 67.2 & 71.3 | 12.5 | 12.3 | 4 & 36 | 24.5 | 39.5 & 53.9 \\
    6     & 700   & 89 | 11 & 35.3 | 64.7 & 71.3 | 10 | 13.6 | 5.1 & 33.3 | 39 | 27.7 & 53.1 \\
  \end{tabular}

  \vspace{1em}
  \textbf{Proportion table for Average method}
  \vspace{0.5em}
  \begin{tabular}{ccccccc}
    \textbf{Intensity} & \textbf{Count} & \textbf{Diabetes} & \textbf{Gender} & \textbf{Ethnicity} & \textbf{Age Group} & \textbf{Age} \\
    group  &       & No | Yes & Male | Female & White | Hispanic | Black | Others & $\le$ 44 | 45-64 | $\ge$ 65 &     \\
    1     & 66    & 92.4 | 7.6 & 45.5 | 54.5 & 71.2 | 13.6 | 10.6 | 4.5 & 69.7 | 21.2 | 9.1 & 39.4 \\ 
    2     & 69    & 84.1 | 15.9 & 47.8 | 52.2 & 72.5 | 10.1 | 15.9 | 1.4 & 59.4 | 15.9 | 24.6 & 46.7 \\ 
    3     & 238   & 79 | 21 & 46.6 | 53.4 & 83.6 | 4.6 | 9.7 | 2.1 & 13.4 | 16.8 | 69.7 & 67.8 \\ 
    4     & 274   & 80.3 | 19.7 & 39.8 | 60.2 & 79.2 | 8 | 10.2 | 2.6 & 23.4 | 18.6 | 58 & 62.6 \\ 
    5     & 276   & 83.3 | 16.7 & 32.2 | 67.8 & 70.3 | 10.9 | 14.9 | 4 & 37.3 | 27.9 | 34.8 & 52.5 \\ 
    6     & 345   & 89.9 | 10.1 & 37.4 | 62.6 & 76.8 | 8.1 | 10.4 | 4.6 & 29.9 | 26.4 | 43.8 & 57.8 \\ 
    7     & 408   & 85.5 | 14.5 & 32.4 | 67.6 & 69.6 | 12.7 | 12.7 | 4.9 & 36.5 | 25 | 38.5 & 53.5 \\ 
    8     & 263   & 85.2 | 14.8 & 44.1 | 55.9 & 69.6 | 10.3 | 15.6 | 4.6 & 27.4 | 38.4 | 34.2 & 55.7 \\ 
    9     & 257   & 91.4 | 8.6 & 28.8 | 71.2 & 67.7 | 12.1 | 14.4 | 5.8 & 42.8 | 44.4 | 12.8 & 47.2 \\ 
  \end{tabular}

  \vspace{1em}
  \textbf{Proportion table for Naive method}
  \vspace{0.5em}
  \begin{tabular}{ccccccc}
    \textbf{Intensity} & \textbf{Count} & \textbf{Diabetes} & \textbf{Gender} & \textbf{Ethnicity} & \textbf{Age Group} & \textbf{Age} \\
    group  &       & No | Yes & Male | Female & White | Hispanic | Black | Others & $\le$ 44 | 45-64 | $\ge$ 65 &     \\
    1     & 153   & 88.9 | 11.1 & 39.2 | 60.8 & 69.3 | 13.7 | 13.1 | 3.9 & 62.7 | 23.5 | 13.7 & 42.3 \\ 
    2     & 253   & 87.4 | 12.6 & 43.9 | 56.1 & 75.9 | 8.7 | 13 | 2.4 & 33.6 | 19.4 | 47 & 57.8 \\ 
    3     & 806   & 82.4 | 17.6 & 33.3 | 66.7 & 72.5 | 8.9 | 13.5 | 5.1 & 30.6 | 27.2 | 42.2 & 56.0 \\ 
    4     & 984   & 86.8 | 13.2 & 39 | 61 & 74.3 | 10.4 | 11.6 | 3.8 & 29.7 | 30.2 | 40.1 & 56.8 \\ 
  \end{tabular}

  \label{prop_nhanes}
\end{table}

\begin{figure}[H]
    \centering
    \includegraphics[width=\linewidth]{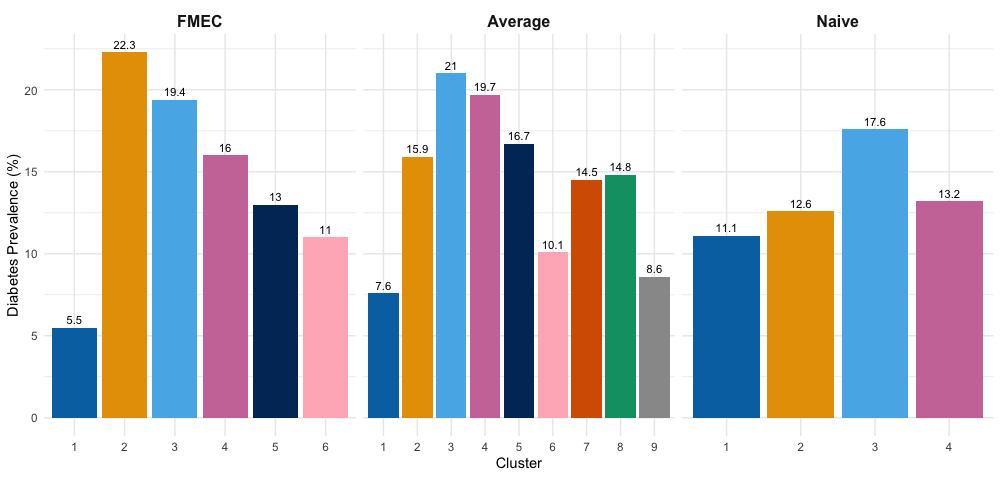}
    \caption{Diabetes prevalence across clusters for different methods}
    \label{prop_diabete_NHANES}
\end{figure}

The distribution of diabetes prevalence across several clusters is shown in Figure ~\ref{prop_diabete_NHANES}. The FMEC method offers the clearest clustering. Cluster 2 had the highest prevalence of diabetes (22.3\%), while cluster 1 has the lowest (5.5\%), possibly due to the small cluster size. It illustrates how well the FMEC method distinguish across groups according to the prevalence of diabetes. Using the FMEC method, the prevalence of diabetes continuously declines as physical activity increases. With diabetes prevalence ranging from 7.6\% in cluster 1 to 21\% in cluster 3, the Average method's clustering offers a little more variety. Nonetheless, similar diabetes rates are seen in several clusters, including clusters 2 and 5, clusters 7 and 8, and others. According to the Naive method, the percentage of people with diabetes varies comparatively little between clusters, ranging from 11.1\% in cluster 1 to 17.6\% in cluster 3.

Clusters derived from the different methods were utilized to perform a logistic regression analysis, with diabetes serving as the outcome variable. Gender, ethnicity, age group, and blood pressure were included as confounding variables. Male participants were treated as the reference group for gender. Ethnicity was categorized into four groups: Non-Hispanic White, Non-Hispanic Black, Hispanic, and Others, with Non-Hispanic White serving as the reference category. Age was grouped into three categories: $\leq 44$, 45–64, and $\geq 65$, with individuals aged $\leq 44$ designated as the reference group. The largest physical activity intensity group was set as the reference group. 

The logistic regression results for the three clustering methods (FMEC, Average and Naive) are presented in Table ~\ref{final_NHANES}. Gender was not significant in any of the models (\(p > 0.3\)), suggesting no substantial difference in diabete prevalence based on gender. Ethnic groups, however, showed consistent and significant associations across all methods. Non-Hispanic Black had the strongest positive association with the diabetes prevalence, with coefficients around 1.1 and \(p < 0.001\) for all methods. Hispanic group also exhibited a significant positive association (\(p < 0.05\)), with smaller coefficients ranging from 0.567 to 0.579. Both age groups, 45–64 and \(\geq\)65 years significantly indicated a higher risk of getting diabetes in older age groups. FMEC and Average methods produced slightly larger coefficients for the 45–64 group (around 2.0) compared to the Naive  method (1.962). The coefficients for intensity groups in the FMEC model were generally positive except group 1, reflecting higher risk of diabetes for lower intensity groups compared to the reference (largest physical intensity). Intensity group 2 showed the strongest association (\(0.858, p < 0.001\)) with diabetes, representing when physical activity decrease, chance of getting diabetes increases. Average method consistently had positive coefficients for all intensity groups. The Naive method included only four intensity groups, with the largest group as the reference. Group 3 showed a significant positive association (\(0.361, p = 0.009\)) between lack of physical activity and risk of diabetes. The Adjusted Akaike Information Criterion (aAIC) values indicate that the FMEC method achieved the best model fit (\(aAIC = 1658.7\)).

\begin{table}[H]
\centering
\caption{Comparison of logistic regression results for FMEC, Average, and Naive Methods with the Largest Group as Reference}
\begin{tabular}{lcccccc}
\toprule
\textbf{Covariate} & \multicolumn{2}{c}{\textbf{FMEC}} & \multicolumn{2}{c}{\textbf{Average}} & \multicolumn{2}{c}{\textbf{Naive}} \\
\cmidrule(lr){2-3} \cmidrule(lr){4-5} \cmidrule(lr){6-7}
                   & Estimate & P-value & Estimate & P-value & Estimate & P-value \\
\midrule
\textbf{Intercept}          & -3.236   & $<0.001^{***}$  & -3.405   & $<0.001^{***}$  & -3.063   & $<0.001^{***}$  \\
\textbf{Female}             & -0.098   & 0.455           & -0.096   & 0.461           & -0.130   & 0.317           \\
\textbf{Hispanic}           & 0.579    & 0.015$^*$       & 0.567    & 0.017$^*$       & 0.567    & 0.016$^*$       \\
\textbf{Non-Hispanic Black} & 1.102    & $<0.001^{***}$  & 1.107    & $<0.001^{***}$  & 1.081    & $<0.001^{***}$  \\
\textbf{Other Race}         & 0.166    & 0.660           & 0.199    & 0.599           & 0.122    & 0.747           \\
\textbf{Age 45-64 (ref: $\leq 44$)} & 1.997 & $<0.001^{***}$  & 2.006 & $<0.001^{***}$  & 1.962 & $<0.001^{***}$  \\
\textbf{Age $\geq 65$ (ref: $\leq 44$)} & 2.213 & $<0.001^{***}$  & 2.258 & $<0.001^{***}$  & 2.324 & $<0.001^{***}$  \\
\textbf{Blood Pressure}     & -0.007   & 0.078$^.$       & -0.007   & 0.073$^.$       & -0.007   & 0.072$^.$       \\
\textbf{Intensity group 1}  & -0.116   & 0.833           & 0.380    & 0.486           & 0.404    & 0.167           \\
\textbf{Intensity group 2}  & 0.858    & $<0.001^{***}$  & 1.057    & 0.013$^*$       & -0.018   & 0.935           \\
\textbf{Intensity group 3}  & 0.539    & 0.003$^{**}$    & 0.722    & 0.014$^*$       & 0.361    & 0.009$^{**}$    \\
\textbf{Intensity group 4}  & 0.496    & 0.021$^*$       & 0.754    & 0.009$^{**}$    & (ref)       & --              \\
\textbf{Intensity group 5}  & 0.245    & 0.188           & 0.710    & 0.015$^*$       & --       & --              \\
\textbf{Intensity group 6}  & (ref)       & --              & 0.044    & 0.882           & --       & --              \\
\textbf{Intensity group 7}  & --       & --              & 0.518    & 0.062$^.$       & --       & --              \\
\textbf{Intensity group 8}  & --       & --              & 0.395    & 0.180           & --       & --              \\
\textbf{Intensity Group 9}  & --       & --              & (ref)    &            & --       & --              \\
\midrule
\textbf{aAIC}               & 1658.9   & --              & 1664.8   & --              & 1666.1 & --              \\
\bottomrule
\end{tabular}
\label{final_NHANES}
\footnotesize 
  \parbox[b]{\linewidth}{\scriptsize Note: AIC (Akaike Information Criterion) is a measure of model fit that accounts for both the goodness of fit and model complexity. Lower AIC values indicate a better balance between model fit and complexity. aAIC (Adjusted AIC) is a modified version of the AIC that accounts for the finite sample size in a model. \\
  Significant level: $<$0.001 ‘***’ 0.001 ‘**’ 0.01 ‘*’ 0.05 ‘.’ 0.1}
\end{table}

\subsection{Physical activity energy expenditure of kids data}
We also applied the proposed FMEC model to the stand-biased desks study (a cluster randomized trial) conducted in Texas is the basis for this practical data application [\cite{benden2014evaluation}]. During the 2012-2013 academic year, 24 teachers from eight elementary schools were recruited and randomly assigned to utilize different types of desks. At baseline, the study included 374 students from second through fourth grades. The student participants were required to wear calibrated BodyMedia SenseWear armband devices during school hours for a week, separately for each semester. These devices recorded individualized step counts and caloric energy expenditure (EE) per minute while being worn. In this paper, we analyzed data at baseline.

The inclusion criteria are: 
\begin{longlist}
    \item[1] Subjects are those who contained five days' data 
    \item[2] Subjects are those who had at least 235 minute level's data during 10:00AM to 13:59PM(240 minutes' interval) each worn day. Missing rate is controlled in order to run the clustering model successfully. 
\end{longlist}

After criteria, the final valid subjects are 195. Since values of energy expenditure data are too small, we multiply with 10 to make the range larger. After getting final sample size, FMEC method, Average method, and Naive method were used to correct EE data with measurement error. Then we use Mclust() package in R to choose the optimal groups. 
Finally a mixed effect model was implemented, with log(BMI) as outcome, the clustered membership as exposure, clustered random effect for teachers nested within schools, and controlling for desk choices, age, gender, ethnicity, to check the effect of functional mixed effect model for clustering when dealing with practical data prone to measurement error.

\begin{figure}[H]
    \centering
    \includegraphics[width=\linewidth]{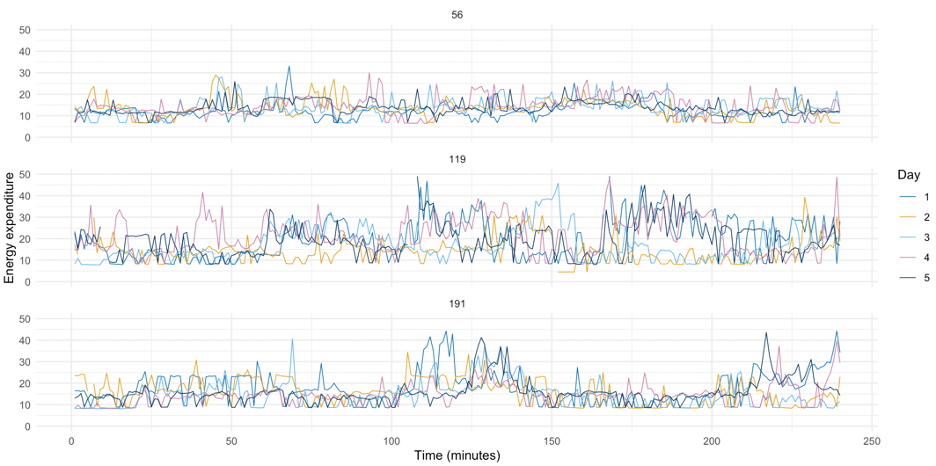}
    \caption{Grouped EE trajectories among different methods}
    \label{trajectory_kids}
\end{figure}

Figure ~\ref{trajectory_kids} illustrates the original EE trajectories for three randomly selected subjects (IDs 56, 119, and 191) over five consecutive days. Each line represents the EE values recorded throughout the day, measured in 1-minute intervals. The trajectories reveal substantial variability in EE patterns across days for each subject. While some peaks and dips are observed consistently, the magnitude and timing of these fluctuations differ by day and subject. For instance, subject 56 exhibits relatively stable EE patterns with sporadic spikes in activity, whereas subject 119 demonstrates more frequent and pronounced variations, particularly toward the end of the day. Subject 191 shows relatively low and steady EE values, with intermittent peaks that suggest bursts of high activity. Across all three subjects, EE values generally range between 0 and 50 units (multipling by 10 with original EE value), with heterogeneity in individual physical activity patterns across days.

The ideal number of clusters was found using the Mclust() program, as seen in Table~\ref{group}. The average energy expenditure (EE) measurements during a four-hour period were then used to assign group memberships in descending order. The average EE line for all of the individuals in a group serves as its defining characteristic. Group 5 has the lowest mean EE value while group 1 has the highest mean EE value over the four-hour period for the FMEC method. If not considering measurement error, depending on the day, the number of clusters based on Mclust() package can range from three to seven groups (day 4 and day 2). This variability demonstrates how Naive measurement influences clustering results and highlights the importance of methods that correct such errors. After applying methods like FMEC or Average method, the optimal number of clusters stabilizes at five. Compared with the Naive method on Day 2 data, these methods merge some subjects into smaller groups while splitting others into more granular clusters. Although the Average and FMEC methods result in the same number of clusters, the distribution of subjects within each cluster differs significantly.

\begin{table}[htbp]
  \centering
  \caption{Group tables among different methods}
    \begin{tabular}{lrrrrrrr}
    FMEC method &       &       &       &       &       &       &  \\
    Group number & 1     & 2     & 3     & 4     & 5     &       &  \\
    Sample size & 54    & 19    & 42    & 25    & 55    &       &  \\
    Average method &       &       &       &       &       &       &  \\
    Group number & 1     & 2     & 3     & 4     & 5     &       &  \\
    Sample size & 37    & 29    & 64    & 25    & 40    &       &  \\
    Naive method (Day 2) &       &       &       &       &       &       &  \\
    Group number & 1     & 2     & 3     & 4     & 5     & 6     & 7 \\
    Sample size & 2     & 35    & 22    & 9     & 52    & 23    & 52 \\
    Naive method (Day 4) &       &       &       &       &       &       &  \\
    Group number & 1     & 2     & 3     &       &       &       &  \\
    Sample size & 64    & 31    & 100   &       &       &       &  \\
    \end{tabular}%
  \label{group}%
\end{table}

\begin{figure}[H]
    \centering
    \includegraphics[width=\linewidth]{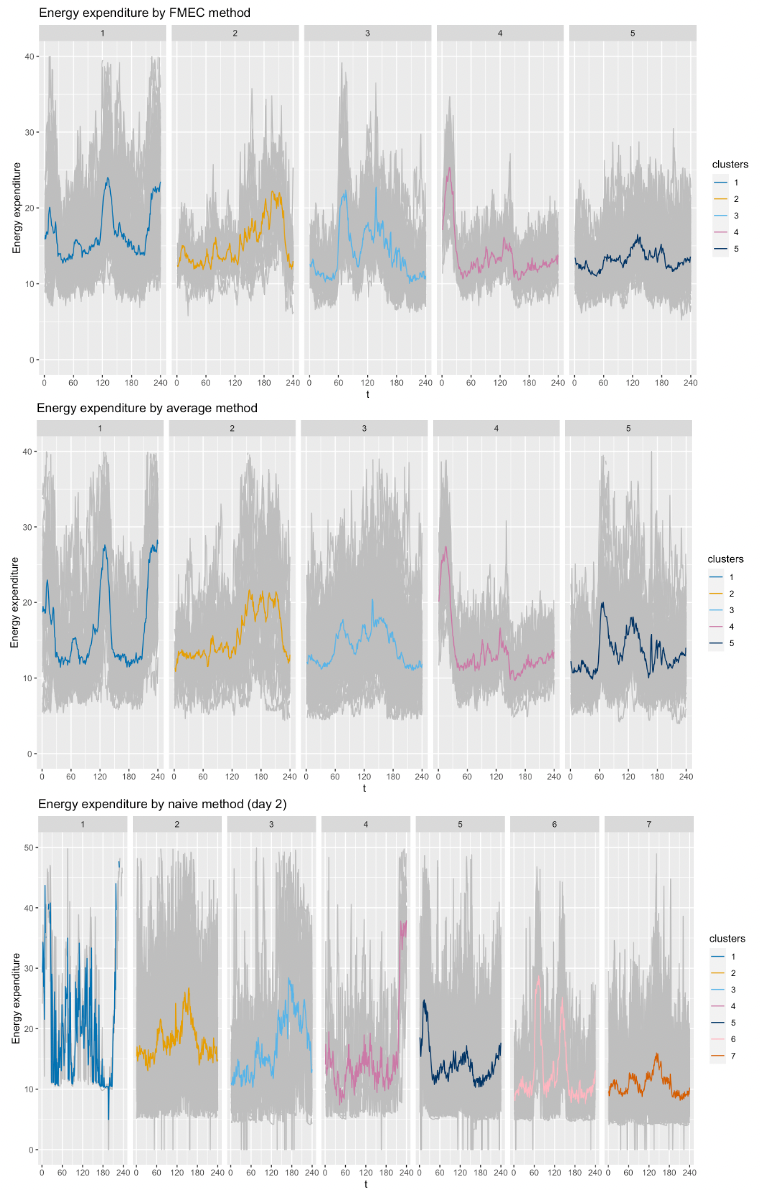}
    \caption{Grouped EE trajectories among different methods}
    \label{group_kids}
\end{figure}

The Figure~\ref{group_kids} indicates trajectories for each cluster by different clustering method. Gray lines are subject-specific trajectories and colorful trajectories are mean-value for each time point. In the FMEC subplot, the mean trajectories closely resemble those observed in the average plot in general, but have some differences. The first group exhibits relatively milder peaks. Group 2 shows an ascending trend over the time period. Group 3 demonstrates more significant fluctuations during the middle two hours. Group 5 displays a consistently smooth and moderate EE intensity. 
In the Average method subplot, group 1 demonstrates a more intensive energy expenditure (EE) range at the initial, middle, and final time points. Group 2 exhibits stable relative EE values after the first two-hour period. Group 3 displays fluctuations in EE during the middle two hours. Group 4 shows a pronounced peak in EE at the beginning, followed by more moderate values. Group 5 experiences varying fluctuations in EE throughout the entire duration.
The third subplot is the Naive method. Here we only plot 7 clusters for day 2 data. Since we randomly choose one day for energy expenditure without measurement error correction, the mean-trajectory data is messy even after clustering. The optimal number of clusters of both FMEC and Average method is 5. While gray lines of naive method are not identified, subject trajectories of average and FMEC method have more distinguished trajectories.

The findings of mixed-effects models comparing three clustering models—FMEC, Average, and Naive method, with BMI as the outcome are shown in Table~\ref{final_kids}. Covariates for treatment group, age, gender, ethnicity, and energy expenditure (EE) levels are included in the models. Women are regarded as the reference group for gender. Non-Hispanic White is the reference category for ethnicity, which also includes Hispanic, Non-Hispanic Black, and Others. The reference group of EE is the one with the highest energy expenditure. Out of the three methods, FMEC exhibits the best model fit with the lowest aAIC score (-118.95), followed by the Average technique (-99.06) and the Naive method (-114.92). With respect to the fixed factors, Non-Hispanic Black ethnicity shows consistently significant higher BMI than the reference group (Non-Hispanic White). Three clusters statistically showing when energy expenditure decreases, BMI decreases for the FMEC method. Less power in identifying significant association is demonstrated by the bigger SEs for Naive method compared with FMEC method.

The variance of the random intercepts for the nested teacher and school levels is zero for the Naive method's random effects. On the other hand, the variance components for random intercepts are not zero for the FMEC and Average approaches. At the school level, the FMEC method's standard deviation (0.035) is marginally higher than the Average method's (0.03). When compared to the Naive technique, the FMEC and Average methods both perform better at capturing the hierarchical variability in the data. In keeping with the original multilevel study design, the FMEC technique provides more difference in variability between schools and instructors within schools.

\begin{table}[H]
\centering
\caption{Comparison of Mixed-Effects Model Results for FMEC, Average, and Naive Methods}
\begin{tabular}{lcccccc}
\toprule
\textbf{Covariate} & \multicolumn{2}{c}{\textbf{FMEC}} & \multicolumn{2}{c}{\textbf{Average}} & \multicolumn{2}{c}{\textbf{Naive}} \\
\cmidrule(lr){2-3} \cmidrule(lr){4-5} \cmidrule(lr){6-7}
& Estimate (SE) & P-value & Estimate (SE) & P-value & Estimate (SE) & P-value \\
\midrule
\multicolumn{7}{l}{\textbf{Fixed Effects}} \\
Intercept          & 3.044 (0.191) & $<0.001$ *** & 2.747 (0.150) & $<0.001$ *** & 3.025 (0.173) & $<0.001$ *** \\
Treatment          & 0.007 (0.025) & 0.759        & -0.006 (0.025) & 0.803       & 0.035 (0.022) & 0.117        \\
Age                & -0.007 (0.017) & 0.693       & 0.014 (0.013) & 0.308       & 0.018 (0.013) & 0.176        \\
Male               & -0.025 (0.022) & 0.259       & -0.009 (0.022) & 0.665       & -0.028 (0.022) & 0.188        \\
Hispanic           & -0.048 (0.038) & 0.207       & -0.038 (0.040) & 0.345       & -0.040 (0.037) & 0.285        \\
Black              & 0.087 (0.037) & 0.020 *      & 0.091 (0.038) & 0.021 *     & 0.085 (0.036) & 0.021 *      \\
Others             & 0.032 (0.038) & 0.385       & 0.038 (0.039) & 0.332       & 0.029 (0.037) & 0.433        \\
EE.Group1          & (ref) &        & (ref) &        & (ref) &      \\
EE.Group2          & -0.101 (0.052) & 0.081       & 0.018 (0.041) & 0.662       & -0.289 (0.104) & 0.006 **     \\
EE.Group3          & -0.143 (0.047) & 0.008 **    & -0.039 (0.034) & 0.278      & -0.317 (0.106) & 0.003 **     \\
EE.Group4          & -0.144 (0.046) & 0.008 **    & -0.061 (0.042) & 0.185      & -0.307 (0.112) & 0.007 **     \\
EE.Group5          & -0.151 (0.031) & $<0.001$ *** & 0.081 (0.037) & 0.041 *     & -0.366 (0.103) & 0.001 ***    \\
EE.Group6          & --             & --          & --             & --         & -0.378 (0.107) & $<0.001$ *** \\
EE.Group7          & --             & --          & --             & --         & -0.440 (0.104) & $<0.001$ *** \\
\midrule
\multicolumn{7}{l}{\textbf{Random Effects (Variance / Std. Dev.)}} \\
Teacher:School (Intercept) & 0.00061 (0.025) & -- & 0.00023 (0.015) & -- & 0 (0) & -- \\
School (Intercept)          & 0.00120 (0.035) & -- & 0.00091 (0.030) & -- & 0 (0) & -- \\
Residual                    & 0.020 (0.143) & -- & 0.023 (0.152) & -- & 0.020 (0.143) & -- \\
\midrule
\textbf{aAIC}              & -118.95 & --          & -99.06 & --          & -114.92 & --          \\
\bottomrule
\end{tabular}
\label{final_kids}
\vspace{0.5cm}
\footnotesize 
  \parbox[b]{\linewidth}{\scriptsize Note: Adjusted Akaike Information Criterion (aAIC) is used to compare model fit while accounting for sample size and complexity. Lower values indicate better fit. \\
  Significant level: $<$0.001 ‘***’ 0.001 ‘**’ 0.01 ‘*’ 0.05 ‘.’ 0.1}
\end{table}

\begin{figure}[H]
    \centering
    \includegraphics[width=\linewidth]{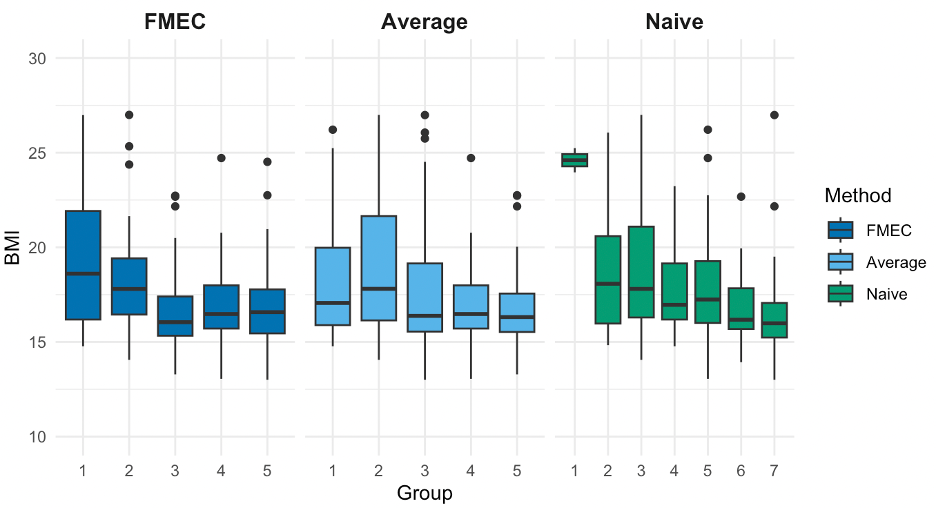}
    \caption{BIM range among different groups and methods}
    \label{bmi_kids}
\end{figure}

The distribution of BMI across various energy expenditure (EE) groups is shown in Figure~\ref{bmi_kids}. The trend shows that the median BMI tends to decline as energy expenditure declines across a variety of groups and techniques. This finding is in line with the regression results. Although this result is statistically consistent, it seems contradictory because a lower BMI is typically assumed to be associated with higher energy expenditure since increased physical activity results in a greater expenditure of calories. This association could be explained by a number of reasons. First, it is important to note that the maximum BMI among subjects was only 26.99. According to established guidelines, a BMI of 25 indicates overweight and 30 indicates obesity in children[\cite{dietz1999introduction}]. Thus, the subject with the highest BMI was only categorized as overweight, with 95\% of subjects having a BMI below 25. Second, the subjects in this study are children aged 9–13, usually in a developmental stage. During this period, energy expenditure patterns can differ widely depending on growth spurts, metabolic changes, and puberty status[\cite{cheng2016energy,pfeiffer2006physical}]. Children with higher BMI may naturally expend more energy due to the increased metabolic cost, even if their physical activity levels are lower[\cite{treuth1998energy}]. Third, the observed energy expenditure might not accurately reflect levels of physical activity. It could be influenced by factors including genetics, sleep patterns, and diet [\cite{ravussin1989relationship,markwald2013impact,vujovic2022late}]. Despite participating in less physical activity, children who have lower energy expenditure may also have lower calorie intake, which results in a lower BMI [\cite{thivel2013daily}].

\section{Discussion}
In order to overcome the difficulties caused by measurement errors in data analysis, we proposed a pairwise mixed effect model (FMEC) in this manuscript. The clustering results can be severely impacted by measurement errors, especially in situations when sample sizes are small or error is high. Our approach is to increase robustness and reliability in these kinds of situations. To assess the FMEC method's performance in various settings, simulations show that FMEC operates consistently effectively, showing particular benefits in situations with high measurement errors and small sample size. However, FMEC is less effective in compound symmetry correlation. The FMEC model fits the data better than the Average and Naive methods, particularly in datasets with small sample sizes, according to the application of the FMEC method to actual datasets. The FMEC method produced better clustering results and fit the data better than the Average and Naive approaches. The use of FMEC in a variety of circumstances and with more correlation structures may be investigated in future research. The method's applicability could be further expanded by expanding it to include multivariate functional responses or dynamic measurement error structures.

\begin{acks}[Acknowledgments]
This research was supported by an award from the National Institutes of Diabetes, Digestive, and Kidney Disease Award number R01DK132385. This research was also supported in part by Lilly Endowment, Inc., through its support for the Indiana University Pervasive Technology Institute.

The author(s) used ChatGPT during the preparation of this work to refine the writing and enhance readability. The content were subsequently reviewed and edited  as necessary and the author(s) take full responsibility for the final publication.
\end{acks}

\bibliographystyle{imsart-nameyear} 
\bibliography{ref}

\end{document}